# Embedding-based Retrieval in Multimodal Content Moderation

Hanzhong Liang*
TikTok
San Jose, CA, USA
hanzhong.liang1@tiktok.com

Jinghao Shi*
TikTok
San Jose, CA, USA
jinghao.shi@tiktok.com

Xiang Shen
TikTok
Bellevue, WA, USA
xiang.shen@tiktok.com

Zixuan Wang
TikTok
San Jose, CA, USA
zixuan.wang1@tiktok.com

Vera Wen
TikTok
San Jose, CA, USA
vera.wen@tiktok.com

Ardalan Mehrani
TikTok
San Jose, CA, USA
ardalan.mehrani@tiktok.com

Zhiqian Chen
TikTok
San Jose, CA, USA
zhiqian.chen1@tiktok.com

Yifan Wu
TikTok
San Jose, CA, USA
yifan.wu@tiktok.com

Zhixin Zhang
TikTok
Singapore
zhangzhixin.01@tiktok.com

## Abstract

Video understanding plays a fundamental role for content moderation on short video platforms, enabling the detection of inappropriate content. While classification remains the dominant approach for content moderation, it often struggles in scenarios requiring rapid and cost-efficient responses, such as trend adaptation and urgent escalations. To address this issue, we introduce an Embedding-Based Retrieval (EBR) method designed to complement traditional classification approaches. We first leverage a Supervised Contrastive Learning (SCL) framework to train a suite of foundation embedding models, including both single-modal and multi-modal architectures. Our models demonstrate superior performance over established contrastive learning methods such as CLIP and MoCo. Building on these embedding models, we design and implement the embedding-based retrieval system that integrates embedding generation and video retrieval to enable efficient and effective trend handling. Comprehensive offline experiments on 25 diverse emerging trends show that EBR improves ROC-AUC from 0.85 to 0.99 and PR-AUC from 0.35 to 0.95. Further online experiments reveal that EBR increases action rates by 10.32% and reduces operational costs by over 80%, while also enhancing interpretability and flexibility compared to classification-based solutions.

## 1 Introduction

In recent years, short video platforms such as TikTok, YouTube Shorts, and Instagram Reels have experienced rapid growth and make video classification an essential task in various features. In particular, video classification[20] models play a key role in content moderation[6] to improve user experience and maintain platform integrity through the detection of inappropriate content. Typically, video classification models predict the category of videos with a sequence of video frames as input. Since textual elements also provide critical context, multimodal classification works such as BLIP [11] and X-VLM [25] integrate textual features with visual representations to enhance performance.

Classification systems often struggle with rapid and cost-efficient responses, particularly in trend adaptation and urgent escalations. A key limitation is their reliance on large-scale datasets, which are often hard to collect for newly emerging risks that require immediate action. Traditional classification models perform poorly in few-shot and zero-shot settings[26], making them ineffective for such scenarios. Moreover, classification models produce only categorical outputs (e.g., class score distributions), limiting their flexibility and granularity when handling high-risk content that demands a more nuanced approach. To address these challenges, we introduce an embedding-based similarity retrieval approach as an alternative. Unlike classification, which learns abstract taxonomic concepts, semantic similarity directly measures risk by comparing content to known high-risk examples. This approach offers

*Both authors contributed equally to this research.

greater adaptability, enabling dynamic updates and real-time risk assessment without retraining the model.

Previous embedding-based approaches, such as CLIP[18] as well as VideoCLIP[22], rely on self-supervised learning framework (SSL), which contrasts different views of the same video to learn similarities between videos. However, visual similarity alone is insufficient for detecting specific trends and risks. To incorporate risk-specific semantics, we adopt supervised contrastive learning (SCL), using risk titles and other fine-grained labels as a semantic proxy for defining similarity relationships. Under this framework, videos sharing the same risk title contain common risk attributes that the model must learn. Conversely, visually similar videos with different risk labels must be distinguished based on label definitions.

Building on these embedding models, we develop the embedding-based retrieval system (EBR), a real-time similarity-driven retrieval system that integrates embedding generation and video retrieval. The EBR system consists of five key components: Seed Selection, Embedding Model, Retrieval Service, Auto-Action Service and Feedback Loop (Figure 1). They work together to detect videos in emerging trends, adapt to distribution shifts, and optimize operations. By continuously refining retrieval results based on real-time feedback, the EBR system enhances precision and adaptability, offering a robust alternative to classification-based approaches for dynamic content moderation and trend detection. We summarize our contributions as follows.

- To the best of our knowledge, we first introduce the embedding-based retrieval approach designed to complement traditional classification methods for content moderation.
- We leverage supervised contrastive learning to develop a suite of single-modal embedding and multimodal embedding models tailored for similarity retrieval task in content moderation.
- We validate the approach through comprehensive offline and online experiments on production data and successfully deploy the model in a real-world production environment.

## 2 Related Works

**Machine Learning based Content Moderation.** As social media platforms expand, the need for effective content moderation continues to grow. Given the vast scale of user-generated content produced daily, machine learning-based automation has become essential. Much work has been done in the past decade to detect hate speech[3], pornography[17], toxic[24], etc. on a variety of signals. Multimodal framework [1, 24] is widely employed as the content on social media contains video, images and text in nature, while few works [16, 23] focus on image moderation or text moderation. With the fast development of multimodal large language model (MLLM), the techniques are introduced to content moderation and achieves decent result [15, 21]. All previous work build a binary classification or multiple classification framework while this work is the first one introduce EBR system into content moderation.

**Embedding-based Retrieval** has been successfully applied in web search[8, 12] or recommendation systems[19] to retrieve items based on user interest or documents based on query text. The techniques has been widely developed and deployed at industrial scale application, e.g. Facebook[8], Snap[19], Walmart[13], and BestBuy[9]. In this work, we first apply this framework in large-scale automated content moderation system.

**Multimodal Representation Learning.** Previous EBR approaches in search engines usually employ a end-to-end two-tower architecture that matches the query and document. However, the EBR system in this paper requires a foundation multimodal model to produce the embeddings of videos and match the seed videos and targeted videos. Contrastive learning approach[2, 7, 18] is widely employed to build a multimodal representation by learning to pull similar pairs together and push dissimilar pairs apart in the embedding space. In addition, the extension to supervised contrastive learning [10] benefits from labeled data and achieves state of the art performance in various downstream tasks.

## 3 Methodology

In this section, we first describe our approach to building embedding models using supervised contrastive learning. Then we introduce the design of the embedding-based retrieval system.

### 3.1 Video Embedding Model

*3.1.1 Training with Supervised Contrastive Learning.* Traditional self-supervised contrastive learning approaches (e.g., SimCLR[2], MoCo[7]) train embedding models by contrasting different views of the same video to learn robust representations. However, these methods lack semantic awareness and may not capture risk-specific distinctions necessary for content moderation. To incorporate granular risk-specific information, we adopt SCL framework[10], which leverages label information to sample positive pairs in addition to augmentations of the same video. This allows the model to learn semantic similarity based on risk attributes, rather than relying solely on visual similarities. Given a multi-viewed batch, the Supervised Contrastive Learning Loss is formulated as follows:

$$L_{sup} = \sum_{i=1}^{2N} \frac{-1}{|P(i)|} \sum_{p \in P(i)} \log \frac{\exp(z_i \cdot z_p/\tau)}{\sum_{a=1, a \neq i}^{2N} \exp(z_i \cdot z_a/\tau)}, \quad (1)$$

where $P(i)$ is the set of indices of all positives in the multiviewed batch distinct from $i$, and $|P(i)|$ is its cardinality. For instance, if $i$ is the index of a video that belongs to the class $y_i$. $P(i)$ will contain the index of all videos having the same class: $y_p = y_i$. $2N$ is the number of augmented samples in the "multiviewed batch". $i = (1..2N)$ the index of an arbitrary augmented sample. The index $i$ is called the anchor. $\tau$ is a scalar temperature parameter.

*3.1.2 Model architectures.* The suite of embedding models we developed includes both single-modal and multimodal models, tailored to different content moderation scenarios: 1) single-modal models are used when visual signals are the primary indicators of risk, 2) multimodal models are employed when critical information comes from multiple modalities, such as text. For single-modal learning, we employ a ViT [4] as the vision encoder, followed by an MLP projection layer to control the feature vector size. In addition, to improve multimodal understanding, a BERT-style text encoder is added to capture textual signals of the video. Additionally, we introduce a cross-attention module before the projection layer, enabling cross-modal interactions and producing a unified multimodal representation.

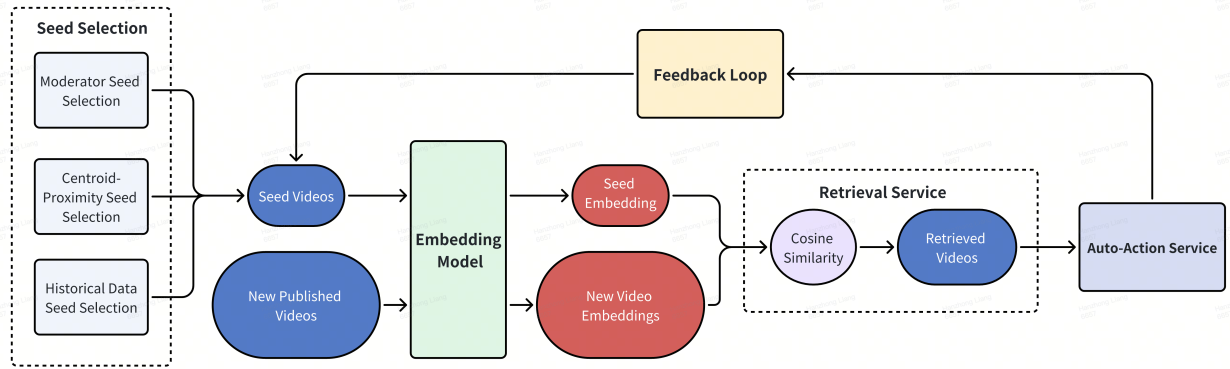

Figure 1: EBR System Design. When a trends emerge, the Seed Selection module selects seed videos, which are then processed by Embedding Models to generate embeddings. The Retrieval Service computes pairwise similarity while the Auto-Action service uses similarity scores as a measure of risk level, triggering different types of actions based on predefined similarity thresholds. The Feedback Loop continuously monitors retrieval quality, adjusting seed videos in real-time to enhance EBR's precision.

## 3.2 Embedding-based Retrieval System (EBR)

Our embedding-based retrieval System integrates embedding generation, video retrieval, and automated decision-making. It consists of five key components: Seed Selection, Embedding Model, Retrieval Service, Auto-Action service, and Feedback Loop (Figure 1).

When a new trend emerges, representative seed videos are selected in multiple ways, including **Centroid-Proximity Seed Selection**, **Historical Data Seed Selection**, and **Moderator Seed Selection**. The selected seed videos, along with newly published videos, are then processed by the embedding model, which computes their vector representations. Depending on the characteristics of the trend, either single-modal or multimodal embedding model will be selected. Next, in the retrieval service, the system computes the pairwise similarity between seed videos and newly published videos. The Auto-Action service interprets similarity scores as a measure of risk level, triggering different types of actions based on predefined similarity thresholds. These actions may include flagging videos for review, applying content restrictions, or escalating cases for further moderation. EBR also incorporates a feedback loop that actively monitors retrieval quality metrics, such as Top-K precision, and dynamically adjusts seed videos and similarity thresholds based on real-time performance. This iterative refinement enhances both retrieval accuracy and operational efficiency, making EBR a scalable and adaptive solution for real-time content moderation.

In our system, seed selection is the most critical components, which ensures the system identifies high-quality seeds. We provide a detailed discussion of this component below.

*3.2.1 Seed Selection Strategies.* Selecting high-quality seeds is crucial for effective retrieval. Here we describe three primary methods for identifying seed videos in our system.

**Centroid-Proximity Seed Selection**. Trending content often forms dense clusters in the embedding space. Therefore, using a clustering algorithm to identify dense regions is effective for seed selection. We try DBSCAN[5] and K-Means, finding that DBSCAN performs better due to its ability to identify arbitrarily shaped clusters without requiring a predefined number of clusters.

**Historical Data Seed Selection**. We leverage historical performance data from past interval $[t-x, t]$ to assess the precision of candidate seeds under various thresholds. If a seed's precision exceeds a specified threshold, it is deemed "good" and can be used. Formally, if at time interval $[t-x, t]$ the system retrieves a set $R_{[t-x,t]}$ of items (with $|R_{[t-x,t]}| = n_{[t-x,t]}$ items) and $r_{[t-x,t]}$ of these are actually relevant (true positives), then the precision $p_{[t-x,t]}$ is:

$$p_{[t-x,t]} = \frac{r_{[t-x,t]}}{n_{[t-x,t]}}. \quad (3)$$

We assume future performance $p_{t+1} \approx p_{[t-x,t]}$. Thus, if $p_{[t-x,t]}$ exceeds our threshold, we conclude that the seed is sufficiently reliable to be used online.

**Moderator Seed Selection**. In addition to algorithmic approaches, human moderators also contribute "golden seeds" to our system. By monitoring user reports and leveraging their domain expertise, moderators can identify emerging trends quickly and upload relevant seed videos to capture related content.

## 4 Experiments

### 4.1 Embedding Model Training

We conduct the experiments on real-world industrial datasets. We employ 430M videos to train the embedding model, including 230M videos with fine-grained labels. Training commences from a pre-trained CLIP-ViT[18] and RoBERTa[14]. We start with an initial learning rate of 1e-5, which decays over time. The training of embedding models take 20 days on 32 Tesla V100 GPU cluster.

### 4.2 Offline Evaluation

*4.2.1 Experiment Setup.* To evaluate our embedding-based retrieval system, we compare it against a classification model baseline. We construct a evaluation dataset of 25 trends covering diverse categories, with individual trend sizes ranging from 200 to 20K. Additionally, we sample random negative data with positive-to-negative ratios varying from 1:50 to 1:1. All the evaluation dataset is unseen to both the EBR and online classification model. For EBR, we randomly select 5% of videos per trend as seed samples. And every seed recall top 200 candidates.

The baseline model is a multimodal classification model[25], pre-trained and fine-tuned under similar configurations as the multimodal embedding model to ensure fair comparison. In the comparion, we employ both single-modal EBR and multimodal EBR to evaluate the performance of proposed approaches.

*4.2.2 Evaluation Metrics.* To provide a comprehensive comparison, we evaluate ROC-AUC, PR-AUC, and F1 scores. For EBR evaluation, each video's classification score is its maximum cosine similarity among all seed videos. We then compute the metrics following the same procedure as the classification baseline. Additionally, we use P@200 to assess retrieval performance among the top-ranked candidates.

*4.2.3 Results.* We draw two main conclusions from Table 1
- EBR significantly outperform the classification baseline across all metrics, demonstrating its superior ability to retrieve emerging inappropriate videos. Online classification model performs poorly on unseen harmful videos.
- In addition, multimodal EBR outperforms single-modal EBR, highlighting the value of multimodal representations for improved retrieval performance.

| Model | P@200 | PR-AUC | ROC-AUC | F1 |
|---|---|---|---|---|
| Baseline | - | 0.3501 | 0.8527 | 0.4047 |
| Single-modal EBR | 0.7972 | 0.8922 | 0.9617 | 0.8693 |
| Multimodal EBR | **0.8351** | **0.9551** | **0.9960** | **0.9450** |

Table 1: Performance metrics across different models. Baseline is a multimodal classification model. Single-modal EBR refers to Embedding based Retrieval system with visual model. Multimodal EBR refers to Embedding based Retrieval system with multimodal model.

## 4.3 Ablation Study

In this section, we analyze the impact of SCL on model training and the effect of seed quantity in the EBR system.

*4.3.1 Supervised Contrastive Learning.* We train four models using contrastive learning on the same dataset: SCL, ResNet50, CLIP, and MoCo V3, and evaluate P@200 across four representative content moderation tasks. As shown in Table 2, SCL outperforms the cross-entropy-based ResNet50 and the self-supervised methods CLIP and MoCo V3 with significant margins.

| Task | CLIP | MoCo | ResNet | SCL |
|---|---|---|---|---|
| Task 1 | 0.390 | 0.748 | 0.831 | **0.868** |
| Task 2 | 0.564 | 0.819 | 0.868 | **0.894** |
| Task 3 | 0.174 | 0.762 | 0.825 | **0.833** |
| Task 4 | 0.091 | 0.654 | 0.679 | **0.693** |

Table 2: Performance comparison of different models across tasks.

*4.3.2 Seed numbers.* We further experiment with EBR using different seed amounts. Figure 2 presents EBR's PR-AUC and F1 scores across various seed percentages. Performance improves with more seeds up to 10%, after which gains become marginal. This suggests that a 5%-10% seed range is sufficient to provide a strong trend representation while maintaining optimal performance in EBR.

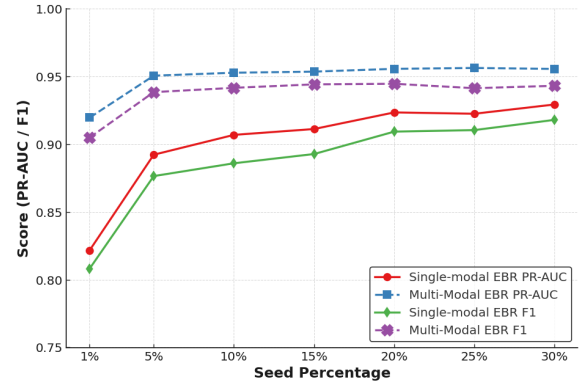

Figure 2: EBR System with different seed percentage.

## 4.4 Online Experiments

*4.4.1 Performance Comparison.* We deploy our EBR system online alongside a classification-model baseline. After incorporating EBR, the total action volume increases by **10.32%** without significant change of other key metrics, demonstrating its effectiveness in content moderation.

To further compare the performances, we also manually evaluate three representative trends separately. Results show that EBR achieves over 95% recall, while the classification model remains below 3%, highlighting EBR's superior capability in trend moderation. These findings align with offline observations, confirming EBR's ability to recall additional trend cases and improve content moderation.

| Trend | Total Volume | EBR Recall | Model Recall |
|---|---|---|---|
| Trend 1 | 141 | 97.87% | 2.13% |
| Trend 2 | 120 | 95.83% | 1.67% |
| Trend 3 | 225 | 100.00% | 0.89% |

Table 3: Comparison of EBR and Model on online trends.

*4.4.2 Cost comparison.* Unlike classification-based solutions that require extensive labeled data, our proposed EBR system operates in a train-free manner. By leveraging only a few seed examples, EBR can respond in real time and handle new trends within one day—significantly faster than traditional model iteration process, which typically takes an average of five days. Our EBR reduces operational costs by over 80% in trend handling. In addition, the proposed approach could be operated by product operators and could greatly save the engineering labor in trend handling.

## 5 Conclusion

In this paper, we introduced a novel embedding-based retrieval (EBR) as a complementary approach to the traditional classification model for content moderation. Our EBR system can be used in production settings as a flexible hot-fixing approach to mitigate immediate harmful content trends. Experiments on real-world datasets, including offline and online, consolidate the effectiveness of EBR. This system not only improves the operation performance, but also enhance interpretability and flexibility compared to classification-based solutions. It has been successfully integrated into production systems with multiple downstream business applications.

## 6 Presenter Bio

Hanzhong Liang is a Machine Learning Engineer at TikTok. He received his M.S. in Electrical and Computer Engineering in Cornell University and B.S in Automation from Shandong University. His research interests lie in retrieval system and large language models.